# Sub-diffraction-limit Observation Realized by Nonlinear Metamaterial Lens

Zhiyu Wang[1,2], Yu Luo[1,2], Liang Peng[1], Jiangtao Huangfu[1], Tao Jiang[1,2], Dongxing Wang[3], Hongsheng Chen[2], Lixin Ran[1]

1. *Department of Information and Electronic Engineering, Zhejiang University, Hangzhou 310027*

2. *The Electromagnetic Academy at Zhejiang University, Zhejiang University, Hangzhou 310027*

3. *School of Engineering and Applied Sciences, Harvard University, Cambridge, MA 02138*

## Abstract

In this paper, we show by experiment that by covering a thin flat nonlinear lens on the sources, the sub-diffraction-limit observation can be achieved by measuring either the near-field distribution or the far-field radiation of the sources at the harmonic frequencies and calculating the inverse Fourier transformation to obtain the sub-wavelength imaging. Especially, the sub-wavelength image calculated from measured far-field data demonstrates very clear resolution. Since metamaterials included with active elements can easily behave strong nonlinearity under very weak incident electromagnetic powers, the application of the nonlinear lens proposed in this paper would have important potential in improving the sub-wavelength resolution in the near future.

The diffraction limit of electromagnetic (EM) wave limits the resolution of detector to the order of the operating wavelength of microwave or light used to distinguish objects [1]. Since the evanescent waves carrying sub-wavelength information about the objects attenuate exponentially in the normal, naturally occurring medium, two adjacent objects closely situated in a distance shorter than the diffraction limit are not able to be distinguished by a "diffraction-limited" system, such as a traditional telescope or microscope. However, by virtue of the artificial metamaterial, the diffraction limit has been no longer a constraint. A "super lens" made of the double negative (DNG) metamaterial restores the evanescent waves to realize a near-field sub-wavelength microscopy [1-3], while a "hyper lens" transforms evanescent modes into propagating ones to observe sub-diffraction-limited objects [5–7] in far-field. In addition to the super lens and hyper lens, Z. Zharov *et al* proposed theoretically a new concept of "nonlinear lens", in which they utilized second harmonics generated from a nonlinear metamaterial slab to realize a super lens, breaking through the diffraction limit in a different way [8]. In this paper, we further investigate the nonlinear lens experimentally. We show that by putting a very thin nonlinear metamaterial flat lens before two sub-diffraction-limited objects and observe the objects at the harmonic frequencies, the distinguishability can be significantly improved, and a sub-diffraction-limit observation to the objects from both near- and far-field ranges can be simultaneously obtained. Especially, the sub-wavelength image calculated from measured far-field data demonstrates very clear resolution.

In a diffraction limited system, one approximation of the diffraction limit is $d = \dfrac{\lambda}{2n\sin\alpha}$, where $n$ is the refractive index of the surrounding medium, $\alpha$ is the maximum incident angle that can enter a lens, and $\lambda$ denotes the wavelength corresponding to the frequency $f_0$ of the incident EM wave. Therefore, for a lens placed in free space, the maximum diffraction limit can be simply estimated by $d_{max} = \lambda/2$. In Fig. 1 (a), the contour map of the emitted electric fields by two dipole antennas in free space with an interval $D = \lambda/2$ along the $x$ direction is shown. We see that in the region immediately surrounding the dipoles, known as "reactive zone" generally estimated by $\lambda/2\pi$ (or about $0.2\lambda$), the two dipoles can still be distinguished if the total fields are measured by a detector scanning along the $x$ direction. However, in a distance beyond the reactive zone, the evanescent waves (or reactive waves) that carry sub-wavelength information about the sources have been rapidly vanished, and the detector can only find one peak and lose the distinguishability, which clearly obeys the diffraction limit. In Fig. 1 (b), we put a thin nonlinear slab inside the reactive zone as a lens, for instance at $y = 0.05\lambda$. Since the nonlinear lens is in the reactive zone and senses the reactive fields carrying the sub-wavelength information, the second and higher order harmonics generated from the lens carry the sub-wavelength information either. If a detector working at these harmonic frequencies are used to scanning the corresponding harmonics in the region to the right of the lens, the dipoles can be distinguished even in a place beyond the former reactive zone. This is because the metamaterial slab has strong nonlinear response, which couples the sub-wavelength information carried by the evanescent wave into high harmonic propagating wave. For instance, a slab of metamaterial with quadratic nonlinear response can form an image of the second-harmonic field of the source being opaque at the fundamental frequency. Since at harmonic frequencies, the diffraction limit turns to be $d_{max}/2$, $d_{max}/3$, and etc., the interval D of the dipoles has exceeded the diffraction limits for the harmonics and therefore the objects can be distinguished by a traditional diffraction-limited system again. Fig. 1 (b) illustrates the contour map of the field of the second harmonic ($2f_0$), showing that the diffraction limit no longer holds.

It is difficult to find naturally occurring medium with strong nonlinearity under weak EM incidence, however, this could be easily done by metamaterials. Early in 1999, J. Pendry *et. al.* showed theoretically that enhanced nonlinear electromagnetic properties could arise from metamaterials [9]. Afterwards, multiple nonlinear metamaterials working with different principles have been investigated theoretically and/or experimentally [10]-[13]. In [14]-[18], metamaterials included with microwave diodes have been reported, both demonstrated strong nonlinearity under small EM incidence. In this paper, we will use the metamaterial sample reported in [18] to fabricate the nonlinear lens.

Fig. 2 (a) shows the photograph of the nonlinear lens. The lens is made by printing lots of I-shaped metallic patterns in alignment on both sides of a 1-mm-thick FR4 substrate, whose relative permittivity is around 4.6, and soldering microwave diodes (Infineon's BAT15-03W) on the gaps of the patterns, shown in the bottom insets of Fig. 2 (a). A direct current (DC) source is used to control the bias voltage of the diodes, choosing a strongly nonlinear region of the volt-ampere characteristic curve of the diodes to obtain a strong nonlinearity. The detailed dimensions for each unit cell are $l = h = 6mm$, $g = 1.6mm$, $w_1 = 0.3mm$, $w_2 = 1mm$, and there are 40 and 48 unit cells along the $x$ and $z$ directions, respectively, yielding a 288-mm-long, 240-mm-wide, 1-mm-thick thin flat lens. For an incident electric field polarized along the z direction, electric resonance can be induced by the metallic resonant patterns and, due the existence of the diodes, enhanced nonlinear electric response can be obtained [18]. The measured nonlinearity of the lens is shown in the upper inset of Fig. 2 (a). In the measurement, an incident monochromatic wave with a 10-dBm power at 3$GHz$ is used to illuminate the lens, and the harmonics are measured by another wide-band horn antenna to the other side of the lens. One clearly sees that the second, third and forth order harmonics at 6, 9 and 12$GHz$ exist with high signal-noise ratios (SNRs), showing strong nonlinearity at weak incidences. At the fundamental frequency, i.e., 3$GHz$, the corresponding wavelength in free space is 100$mm$. Comparing with the 6-mm periodicity of the unit cells in the lens, the metamaterial can be regarded as an effective media at least at fundamental and lower order harmonic frequencies.

The experimental setup for the observation of resolution improvement is shown in Fig. 2 (b). Two standard dipole antennas polarized along the z direction driven by equal-amplitude and in-phase monochromatic waves serve as sources, and a third identical dipole antenna serves as a detector. The interval D between the sources is set to be $\lambda/2$ or $\lambda/4$ within the diffraction limit. The lens is put between the sources and the detector, with a distance very close to the sources. The input monochromatic wave is generated by a Vector Signal Generator (Agilent E8267C) and the output spectrum is detected by a Spectrum Analyzer (Advantest R3271A).

The experiments are conducted in a microwave anechoic chamber in both near- and far-field ranges. The frequency and power of the input monochromatic wave is selected to be 3*GHz* and 30*dBm*, respectively. In the near-field measurement, we move the third dipole antenna along x axis at four different distances, i.e., *y*=10*mm*, 20*mm*, 30*mm* and 40*mm*, respectively. In the far-field measurement, the sources as well as the lens are placed in the quiet zone of the chamber and rotated, meanwhile a broadband horn antenna $6\lambda$ (600*mm*) away from the sources is served as a receiver to measure the far-field radiation pattern.

For comparison, we firstly perform a control experiment without the insertion of the lens. When the incidence is at 3*GHz* and the interval D is $\lambda/2$, i.e., 50*mm*, the measured electric fields is shown in Fig. 3 (a)-(c). As expected, we find from the near-field distribution in Fig. 3 (a) and 3 (b) that when the detector is leaving from the sources, the distinguishability to the two sources drastically degrades, and after $y > 20mm$, or about $0.2\lambda$, which is just around the boundary of the reactive zone, the distinguishability completely loses. Similarly, from the far-field pattern in Fig. 3 (c), we can only find one lobe, implying that we cannot recover the position information of the two sources from the far-field data either.

Then we insert the lens at the place where $y = 0.05\lambda$, or 5*mm* away from the sources, and perform the same measurement but at the second-harmonic frequency, i.e., 6*GHz*. The data are shown in Fig. 3 (d)-(f). Compared with the control experiment, we see clearly from Fig. 3 (d) and 3 (e) that the near-field distribution changes obviously, and even when y = 40 mm, we may still observe the variation of the near-field. For far-field, the pattern shown in Fig. 3 (f) now has three lobes, which is also completely different from that in Fig. 3 (c). These clearly indicate that in this case the diffraction limit no longer holds. We will show later that the position information, or image, of the sources can be retrieved from both the near- and far-field data by appropriate algorithm.

Next, while keeping other experimental setup unchanged, we change the interval D to $\lambda/4$. In this case, since the diffraction limit for the second-harmonic frequency is also $\lambda/4$, we may expect that the distinguishability will degrade again if we still measure at the second-harmonic frequency. The measured data are shown in Fig. 3 (g)-(i). We see that both the near-field distribution and far-field pattern return to the shapes in Fig. 3 (a)-(c) and the distinguishability loses again beyond the reactive zone. However, if we measure at the forth-harmonic frequency, i.e., 12*GHz*, in the same circumstance, we again obtain the sub-diffraction-limit observation capability immediately, shown in Fig. 3 (j)-(l).

According to antenna theory, the source field distribution in an aperture can be calculated by performing an inverse Fourier transformation (IFT) to its far-field pattern [19]. In our case, we can calculate the electric field distribution $E(x)$ at y = 0 using a normalized approximate equation

$$E(x) = \int_0^\pi E(\varphi)e^{-j2\pi x\cos\varphi/\lambda_m}d\varphi , \qquad (1)$$

where $E(\varphi)$ is the far-field pattern, $\varphi$ is the azimuth angle and $\lambda_m$ is the wavelength corresponding to the frequency at which the pattern is measured. Applying equation 1 to the patterns in Fig. 3 (c) and 3 (f), we obtain the field distribution, or virtual image, of the source antennas shown in Fig. 4 (a) and 4 (b), respectively. The phase information is not measured, so in the calculation, the side lobe level (SLL) is treated to be negative, knowing that the pattern is caused by two linear antennas [19]. We see that without the lens (Fig. 4 (a)), the image only implies a single source, while with the lens (Fig. 4 (b)), the image clearly illustrates two discrete sources with D=61.5*mm*. Here, the IFT actually acts a role as a virtual "digital" lens, similar with that optical one used in the hyper lens experiment in [7].

To retrieve the image from the near-field data in Fig. 3 (a) and 3 (d), equation 1 no longer holds. However, in the area beyond the reactive zone, the near-field has decayed exponentially, equation 1 still can be used to imprecisely calculate the image, shown in Fig. 4 (c) and 4 (d), respectively. We see that in the case without the lens (Fig. 4 (c)), again the sources can not be distinguished, while with the lens (Fig. 4 (d)), the sub-diffraction-limit imaging appears, although with longer distance *y*, the signal-noise ratio (SNR) of the image is getting worse either.

In conclusion, we show by experiments that by covering a thin flat nonlinear lens on the sources, the sub-diffraction-limit observation can be achieved by measuring either the near-field distribution or the far-field radiation of the sources at the harmonic frequencies and calculating the IFT to obtain the sub-wavelength imaging. The higher order harmonics are used, the higher resolution is obtained. Figure 4(b) shows that the experimental far-field imaging behaves very good performance. Since the far-field pattern keeps its shape unchanged in all the far-field range, the imaging will not be influenced by the distance, as long as the pattern has a good enough SNR. In further research, with the phase information, we can retrieve better results by the similar method in [20]. It has been shown that metamaterials included with active elements can easily behave strong nonlinearity with very higher order harmonics under very weak incident EM powers. Therefore, the application of the nonlinear lens proposed in this paper would have important potential in improving the sub-wavelength resolution in the near future.

Companies, Inc, Chapter 15.2 (2002).

[20] C. Barsi, W. Wan, J. W. Fleischer, "Imaging through nonlinear media using digital holography," Nature Photonics, **3**, 211-215 (2009).

**Figure captions**

Fig. 1. (a) Near-field distribution of two linear sources with a interval D within the diffraction limit. (b) Near-field distribution for the second harmonic with the sources covered by a nonlinear flat lens.

Fig. 2. (a) The nonlinear metamaterial lens and (b) the experimental setup.

Fig. 3. Near- and far-field experimental results.

Fig. 4. IFT of near- and far-field data with 50*mm* interval between two sources. (a) IFT of far-field data at 3*GHz* without lens. (b) IFT of far-field data at 6*GHz* with lens. (c) IFT of near-field data at 3*GHz* without lens. (d) IFT of near-field data at 6*GHz* with lens.

Figure 1.

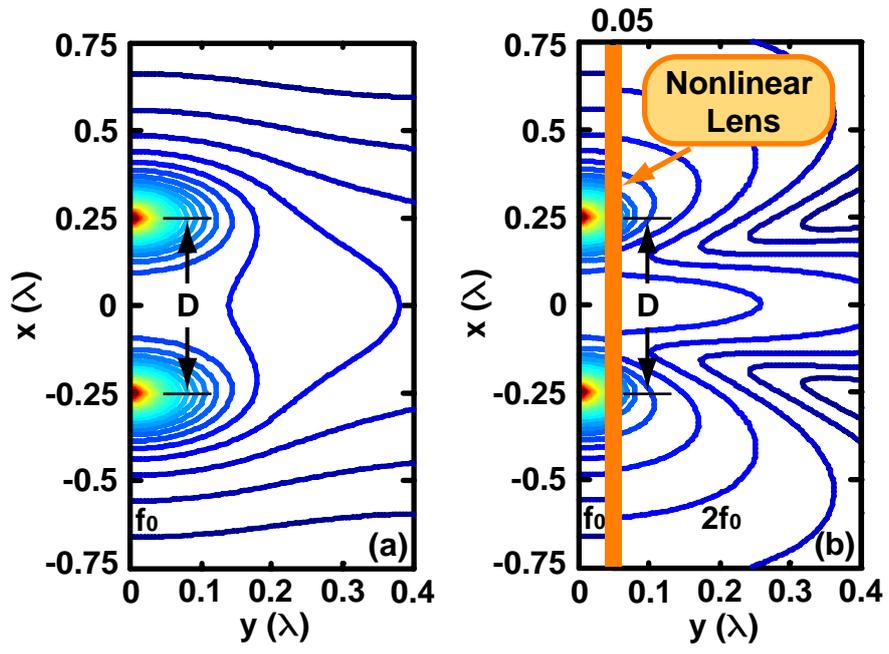

Figure 2.

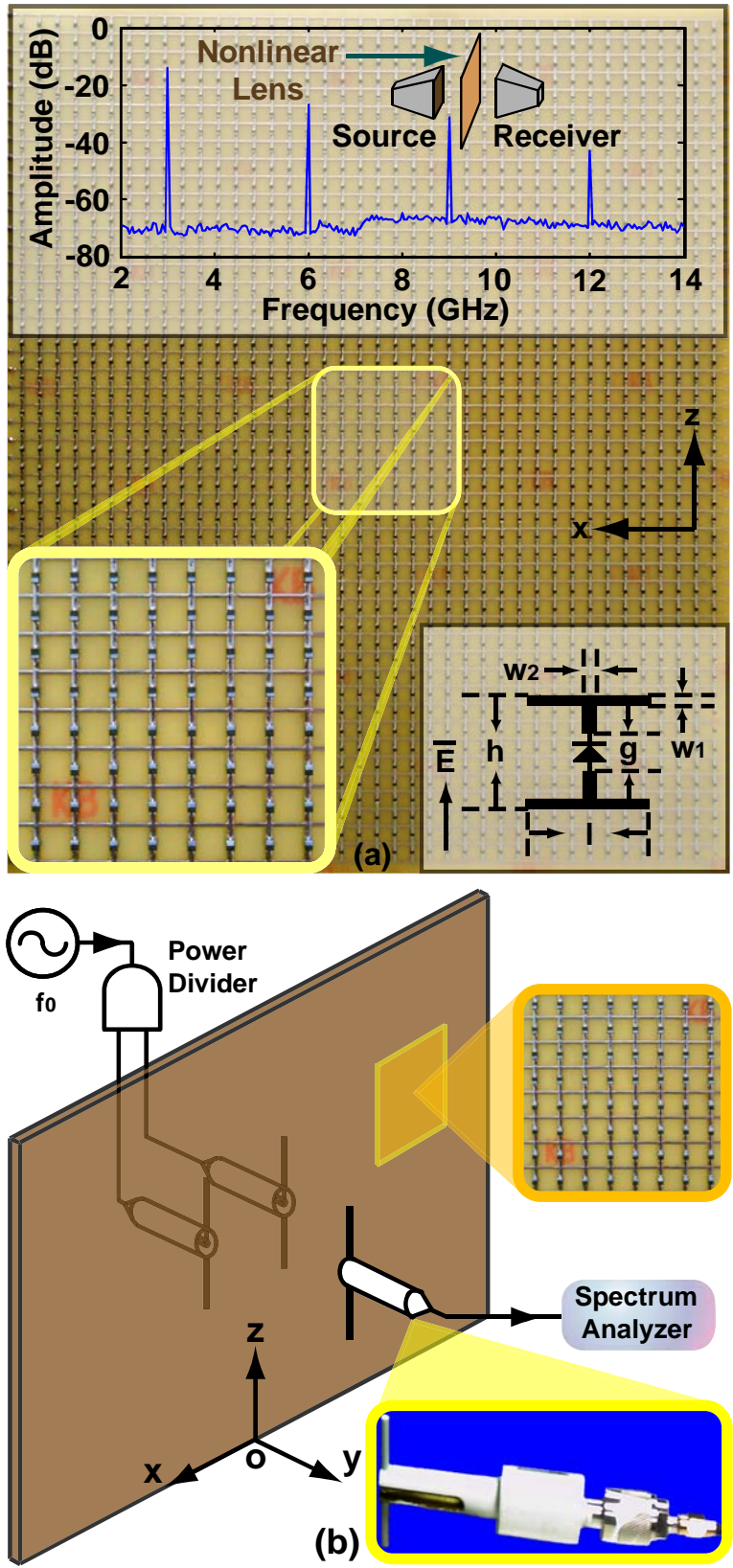

Figure 3.

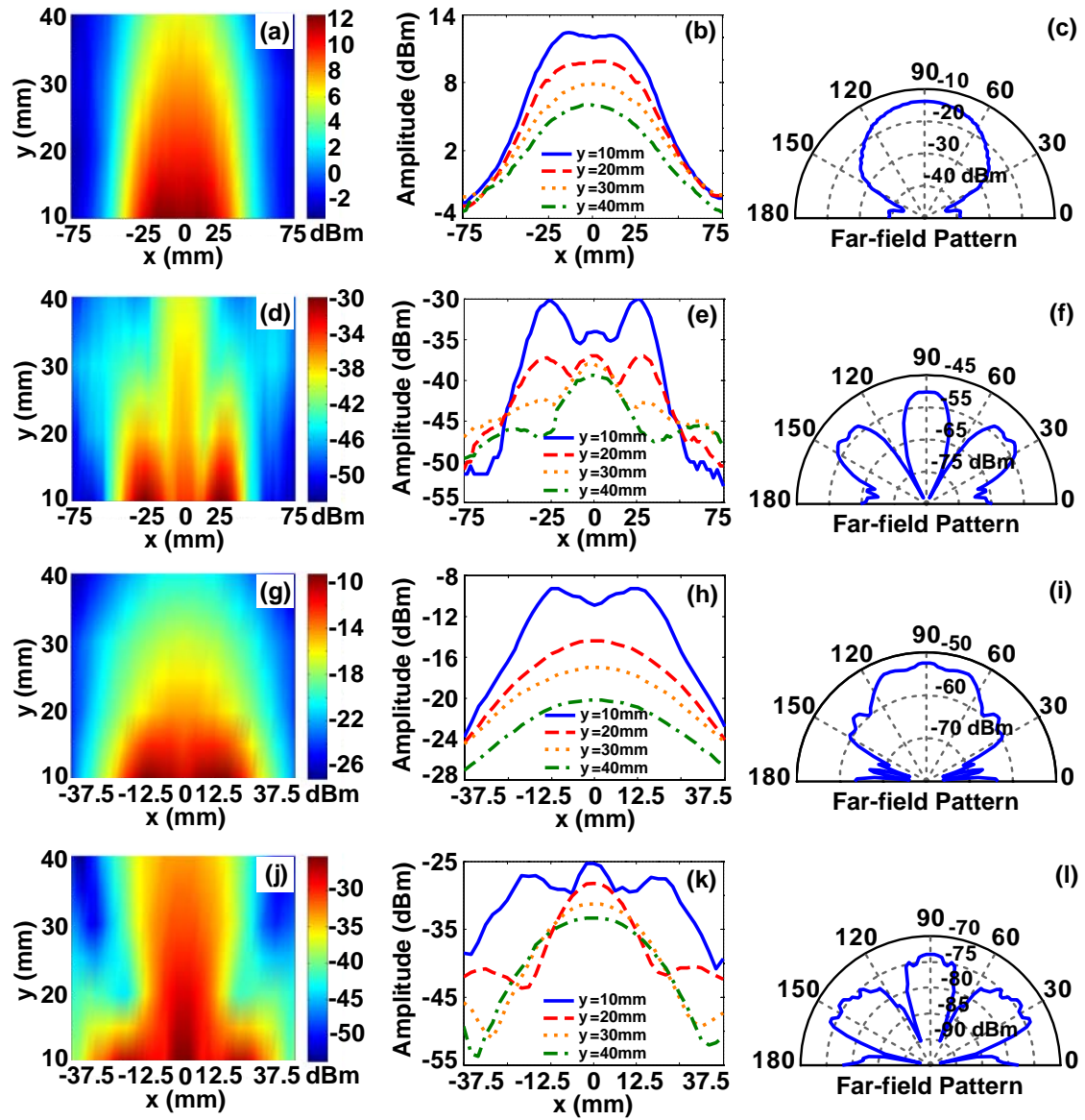

Figure 4.

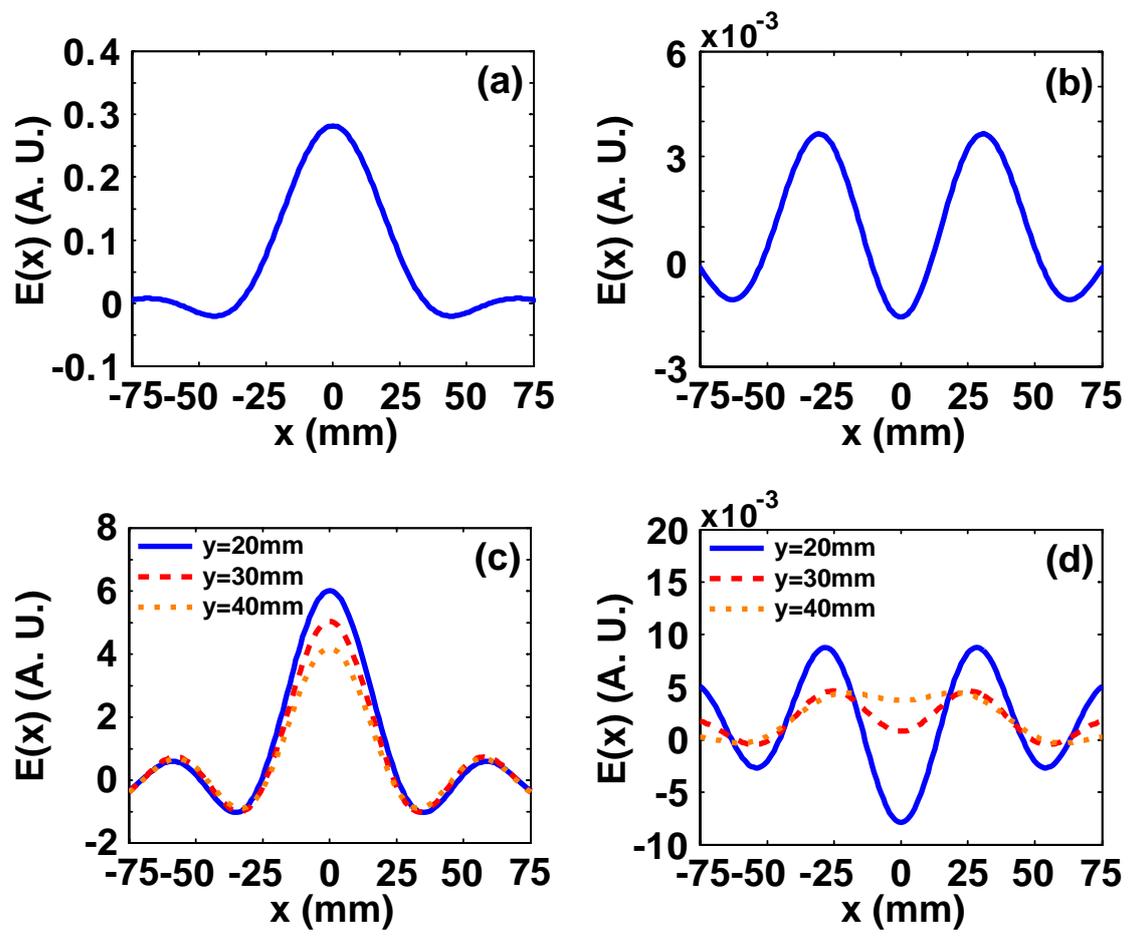